\documentclass[12pt]{article}
\usepackage{graphicx}
\begin{document}

\begin{center}

{\Large \bf Can the quark model be relativistic enough to include the parton model?}

\vspace{3ex}

Y. S. Kim~\footnote{email: yskim@physics.umd.edu}\\
Center for Fundamental Physics, University of Maryland, \\
College Park, Maryland 20742, U.S.A.
\vspace{3ex}

Marilyn E. Noz~\footnote{email: marilyne.noz@gmail.com} \\
Department of Radiology, New York University,\\
New York, New York 10016, U.S.A.

\end{center}

\vspace{3ex}

\begin{abstract}
Since quarks are regarded as the most fundamental particles which
constitute hadrons that we observe in the real world, there are many
theories about how many of them are needed and what quantum numbers
they carry.  Another important question is what keeps them inside the
hadron, which is known to have space-time extension.  Since they are
relativistic objects, how would the hadron appear to observers in
different Lorentz frames?  The hadron moving with speed close to that
of light appears as a collection of Feynman's partons.  In other words,
the same object looks differently to observers in two different frames,
as Einstein's energy-momentum relation takes different forms for those
observers.  In order to explain this, it is necessary to construct a
quantum bound-state picture valid in all Lorentz frames.  It is noted
that Paul A. M. Dirac studied this problem of constructing relativistic
quantum mechanics beginning in 1927.  It is noted further that he
published major papers in this field in 1945, 1949, 1953, and in 1963.
By combining these works by Dirac, it is possible to construct a
Lorentz-covariant theory which can explain hadronic phenomena in the
static and high-speed limits, as well as in between.  It is shown also
that this Lorentz-covariant bound-state picture can explain what we
observe in high-energy laboratories, including the parton distribution
function and the behavior of the proton form factor.
\end{abstract}

\newpage
\section{Introduction}\label{intro}

The hydrogen atom played the pivotal role in the development of
quantum mechanics.  Its discrete energy levels led to the concept
of a localization condition for the probability distribution, and
thus to the bound-state picture of quantum mechanics.  Likewise,
the quark model is still playing the central role in high-energy
physics~\cite{gell64}.  In this model, hadrons are bound states
of more fundamental particles called``quarks'' with their own
internal quantum numbers, such as isospins, unitary spins, and
then flavors.

Thus, the symmetry of combining these quantum numbers has been and
still is an important branch of physics.  Unlike the proton and
electron in the hydrogen atom,  quarks have never been observed
as free particles.  They are always confined inside the hadron.
Thus the only way of determining their properties is through
observing the symmetry properties of hadrons.

Then there comes the question of the binding forces between them,
and the dynamics governing those forces.  If the hadrons are
assumed to be quantum bound states, there are localized probability
distributions whose boundary conditions generate discrete mass
spectra.  This aspect of quantum mechanics is well known.  On
the other hand, it is not yet completely clear how the localized
probability distribution would look to observers in different
Lorentz frames.  Protons coming from high-energy accelerators are
quantum bound states seen in a Lorentz frame moving very fast
with respect to their rest frame.  This is the question on which
we would like to concentrate in this review paper.

There are then three steps.  First, we have to assemble the
physical principles needed to construct this scheme.  We shall
need space-time transformation laws of special relativity and
uncertainty principles of quantum mechanics applicable to position
and momentum variables.  Since we are interested in constructing
a Lorentz-covariant theory, we need the time-energy uncertainty
relation.   However, this time-energy relation does not allow
excited states, and has to be treated differently.  This is the
first hurdle we have to overcome.

The second step is to construct a mathematical formalism which
will accommodate all the physical conditions presented in the
first step.  As always, harmonic oscillators serve as test models
for all new theories.  We shall construct a formalism based on
harmonic oscillators, whose wave functions satisfy Lorentz-covariant
boundary conditions, orthogonality conditions, the difference
between position-momentum and time-energy uncertainty relations.
This covariant oscillator formalism will satisfy all physical laws
of quantum mechanics and special relativity.

The third step is to see whether the theory tells the story of
the real world.  For this purpose, we discuss in detail the proton
form factors and Feynman's parton picture~\cite{fey69a,fey69b}.
Indeed, it has been the most outstanding issue in high energy
physics whether the quark model and parton model are two different
limiting cases of one Lorentz-covariant entity.  We examine this
issue in detail.

This review paper is largely based on the papers published by the
present authors.  But we are not the first ones to approach the
difficult problem of constructing a Lorentz-covariant picture of
quantum bound states.  Indeed, this problem was recognized earlier
by Dirac, Wigner, and Feynman.  We shall present a review of their
valiant efforts in this direction.  These great physicists
constructed big lakes.  We are connecting these lakes to construct
a canal leading to an understanding of relativistic bound states
applicable to the quark model and the parton model.

Einstein was against the Copenhagen interpretation of quantum
mechanics.  Why was he so against it?  The present form of quantum
mechanics is regarded as unsatisfactory because of its probabilistic
interpretation.  At the same time, it is unsatisfactory because
it does not appear to be Lorentz-covariant.  We still do not
know how the hydrogen atom appears to a moving observer.

While relativity was Einstein's main domain of interest, why did he
not complain about the lack of Lorentz covariance?  It is possible
that Einstein was too modest to mention relativity, and instead
concentrated his complaint against its probabilistic interpretation.
It is also possible that Einstein did not want to send his most
valuable physics asset to a battle ground.  We cannot find a definite
answer to this question, but it is gratifying to note that the
present authors are not the first ones to question whether the
Copenhagen school of thought is consistent with the concept of
relativity~\cite{kn06aip}.

Paul A. M. Dirac was never completely happy with the Copenhagen
interpretation of quantum mechanics, but he thought it was a
necessary temporary step.  In that case, he thought we should
examine whether quantum mechanics is consistent with special
relativity.

As for combining quantum mechanics with special relativity, there
was a giant step forward with the construction of the present form of
quantum field theory.
It leads to a Lorentz covariant S-matrix which enables us
to calculate scattering amplitudes using Feynman diagrams.
However, we cannot solve bound-state problems or localized
probability distributions using Feynman diagrams~\cite{fkr71}.

Dirac was never happy with the present form of field
theory~\cite{dir70}, particularly with infinite quantities
in its renormalization processes.  Furthermore, field theory
never addresses the issue of localized probability.  Indeed,
Dirac concentrated his efforts on determining whether localized
probability distribution is consistent with Lorentz covariance.

In 1927~\cite{dir27}, Dirac noted that there is a time-energy
uncertainty relation without time-like excitations.  He pointed
out that this space-time asymmetry causes a difficulty in
combining quantum mechanics with special relativity.

In 1945~\cite{dir45}, Dirac constructed four-dimensional
harmonic oscillator wave functions including the time variable.
His oscillator wave functions took a normalizable Gaussian form,
but he did not attempt to give a physical interpretation to this
mathematical device.

In 1949~\cite{dir49}, Dirac emphasized that the task of building
a relativistic quantum mechanics is equivalent to constructing a
representation of the Poincar\'e group.  He then pointed out
difficulties in constructing such a representation.
He also introduced the light-cone coordinate system.

In 1963~\cite{dir63}, Dirac used two coupled oscillators to
construct a representation of the $O(3,2)$ deSitter group
which later became the basic mathematics for two-photon
coherent states known as squeezed states of light~\cite{knp91}.

In this paper, we combine all of these works by Dirac to make
the present form of uncertainty relations consistent with
special relativity.  Once this task is complete, we can start
examining whether the probability interpretation is ultimately
valid for quantum mechanics.

We then use this Lorentz-covariant model to understand the
relativistic aspect of the quark model.  The most outstanding
problem is whether the quark model and the parton model can
be combined into one Lorentz-covariant model, as in the case of
Einstein's energy-momentum relation $E = \sqrt{m^2 + p^2}$
valid for all values of $(p/m).$

We know that the quark model is valid for small values of $(p/m).$
We consider Feynman's parton picture for the limit of large $(p/m)$.
We then consider the proton proton form factor for $(p/m)$ between
these two limiting cases.

In Sec.~\ref{dirac11}, we review four of Dirac's major papers by
giving graphical illustrations.  Section~\ref{dirac22} is devoted
to combining all four of Dirac's papers into one Lorentz covariant
model for quantum bound states.  In Sec.~\ref{quarkmo}, we discuss
the parton model and the proton form factor to show that the
model is consistent with what we observe in high-energy laboratories.

\section{Dirac's Attempts to make Quantum Mechanics Lorentz
Covariant}\label{dirac11}

Paul A. M. Dirac made it his lifelong effort to formulate quantum
mechanics so that it would be consistent with special relativity.
In this section, we review four of his major papers on this subject.
In each of these papers, Dirac points out fundamental difficulties
in this problem.

In 1927~\cite{dir27}, Dirac notes that there is an uncertainty relation
between the time and energy variables which manifests itself in
emission of photons from atoms.   He notes further that there are no
excitations along the time or energy axis, unlike Heisenberg's
uncertainty relation which allows quantum excitations.  Thus, there is
a serious difficulty in combining these relations in the Lorentz-
covariant world.

In 1945~\cite{dir45}, Dirac considers the four-dimensional harmonic
oscillator and attempts to construct a representation of the Lorentz
group using the oscillator wave functions.  However, he ends up with
the wave functions which do not appear to be Lorentz-covariant.

In 1949~\cite{dir49}, Dirac considers three forms of relativistic
dynamics which can be constructed from the ten generators of the
Poincar\'e group.  He then imposes subsidiary conditions necessitated
by the existing form of quantum mechanics.  In so doing, he ends up
with inconsistencies in all three of the cases he considers.

In 1963~\cite{dir63}, he constructed a representation of the $O(3,2)$
deSitter group using coupled harmonic oscillators.  Using step-up and
step-down operators, he constructs a beautiful algebra, but he makes
no attempts to exploit the physical contents of his algebra.

In spite of the shortcomings mentioned above, it is indeed remarkable
that Dirac worked so tirelessly on this important subject.  We are
interested in combining all of his works to achieve his goal of making
quantum mechanics consistent with special relativity.  Let us
review the contents of these papers in detail, by transforming Dirac's
formulas into geometrical figures.

\begin{figure}
\centerline{\includegraphics[scale=1.2]{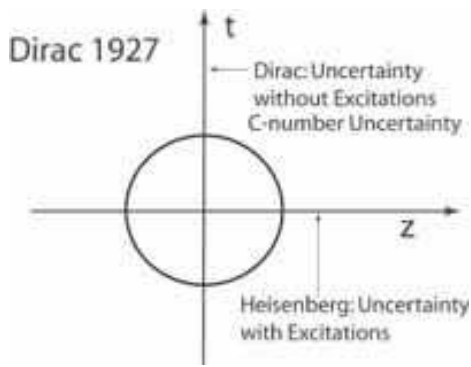}}
\vspace{5mm}
\caption{Space-time picture of quantum mechanics.  There are
quantum excitations along the space-like longitudinal direction,
but there are no excitations along the time-like direction.  The
time-energy relation is a c-number uncertainty
relation.}\label{diracf27}
\end{figure}

\subsection{Dirac's C-Number Time-Energy Uncertainty
Relation}\label{quantu}

Even before Heisenberg formulated
his uncertainty principle in 1927, Dirac studied the uncertainty
relation applicable to the time and energy variables~\cite{dir27,wig72}.
This time-energy uncertainty relation was known before 1927 from the
transition time and line broadening in atomic spectroscopy.  As soon
as Heisenberg formulated his uncertainty relation, Dirac considered
whether the two uncertainty relations could be combined to form a
Lorentz covariant uncertainty relation~\cite{dir27}.

He noted one major difficulty.  There are excitations along the
space-like longitudinal direction starting from the position-momentum
uncertainty, while there are no excitations along the time-like
direction. The time variable is a c-number.  How then can this
space-time asymmetry be made consistent with Lorentz covariance,
where the space and time coordinates are mixed up for moving
observers.

Heisenberg's uncertainty
relation is applicable to space separation variables.  For instance,
the Bohr radius measures the difference between the proton and
electron.  Dirac never addressed the question of the separation in
time variable or the time interval even in his later papers.

As for the space-time asymmetry, Dirac came back to this question
in his 1949 paper~\cite{dir49} where he discusses the ``instant form''
of relativistic dynamics.  He talks indirectly about the possibility
of freezing three of the six parameters of the Lorentz group, and
thus working only with the remaining free parameters.

This idea was presented earlier by Wigner~\cite{wig39,knp86} who
observed that the internal space-time symmetries of particles are
dictated by his little groups with three independent parameters.

\begin{figure}
\centerline{\includegraphics[scale=1.6]{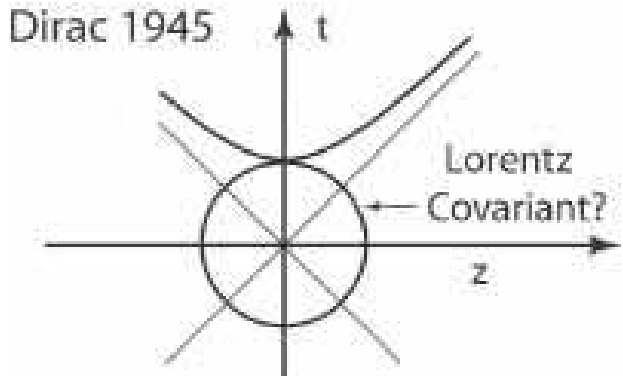}}
\caption{Dirac's four-dimensional oscillators localized in a closed
space-time region.  This is not a Lorentz-invariant concept.  How about
Lorentz covariance?
}\label{diracf45}
\end{figure}

\subsection{Dirac's four-dimensional oscillators}

During World War II, Dirac was looking into the possibility of
constructing representations of the Lorentz group using harmonic
oscillator wave functions~\cite{dir45}.  The Lorentz group is the
language of special relativity, and the present form of quantum
mechanics starts with harmonic oscillators.  Therefore, he was
interested in making quantum mechanics Lorentz-covariant by
constructing representations of the Lorentz group using harmonic
oscillators.

In his 1945 paper~\cite{dir45}, Dirac considers the Gaussian form
\begin{equation}\label{ground4}
\exp\left\{- {1 \over 2}\left(x^2 + y^2 + z^2 + t^2\right)\right\} .
\end{equation}
We note that this Gaussian form is in the $(x,~y,~z,~t)$
coordinate variables.  Thus, if we consider a Lorentz boost along the
$z$ direction, we can drop the $x$ and $y$ variables, and write the
above equation as
\begin{equation}\label{ground}
\exp\left\{- {1 \over 2}\left(z^2 + t^2\right)\right\} .
\end{equation}
This is a strange expression for those who believe in Lorentz
invariance where $\left(z^2 - t^2\right)$ is an invariant quantity.

On the other hand, this expression is consistent with his earlier papers
on the time-energy uncertainty relation~\cite{dir27}.  In those papers,
Dirac observed that there is a time-energy uncertainty relation, while
there are no excitations along the time axis.

Let us look at Fig.~\ref{diracf27} carefully.  This figure is a pictorial
representation of Dirac's Eq.(\ref{ground}),  with localization in both
space and time coordinates.  Then Dirac's fundamental question would be
how to make this figure covariant?  This question is illustrated in
Fig.~\ref{diracf45}.  This is where Dirac stops.  However, this is not
the end of the Dirac story.

\begin{figure}
\centerline{\includegraphics[scale=1.4]{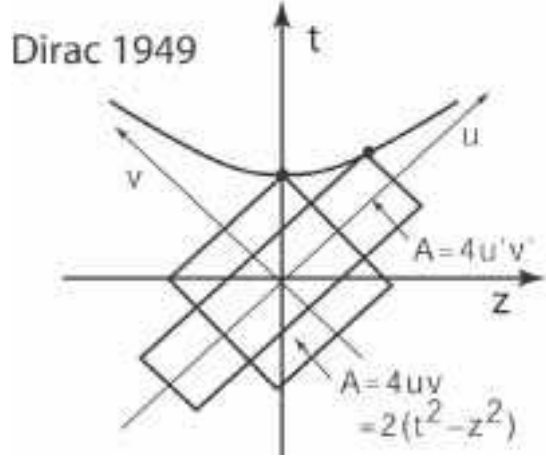}}
\vspace{2mm}
\caption{Lorentz boost in the light-cone coordinate system.  The
boost traces a point along the hyperbola.  The boost also squeezes the
square into a rectangle.}\label{diracf49}
\end{figure}

\subsection{Dirac's light-cone coordinate system}\label{lightcone}

In 1949, the Reviews of Modern Physics published a special issue to
celebrate Einstein's 70th birthday.  This issue contains Dirac paper
entitled ``Forms of Relativistic Dynamics''~\cite{dir49}.
In this paper, he introduced his light-cone coordinate system,
in which a Lorentz boost becomes a squeeze transformation, where one
axis expands while the other contracts in such a way that their product
remains invariant.

When the system is boosted along the $z$ direction, the transformation
takes the form
\begin{equation}\label{boostm}
\pmatrix{z' \cr t'} = \pmatrix{\cosh(\eta/2) & \sinh(\eta/2) \cr
\sinh(\eta/2) & \cosh(\eta/2) } \pmatrix{z \cr t} .
\end{equation}

This is not a rotation, and people still feel strange about this
form of transformation.  In 1949~\cite{dir49}, Dirac introduced his
light-cone
variables defined as~\cite{dir49}
\begin{equation}\label{lcvari}
u = (z + t)/\sqrt{2} , \qquad v = (z - t)/\sqrt{2} ,
\end{equation}
the boost transformation of Eq.(\ref{boostm}) takes the form
\begin{equation}\label{lorensq}
u' = e^{\eta/2 } u , \qquad v' = e^{-\eta/2 } v .
\end{equation}
The $u$ variable becomes expanded while the $v$ variable becomes
contracted, as is illustrated in Fig.~\ref{diracf49}.  Their product
\begin{equation}
uv = {1 \over 2}(z + t)(z - t) = {1 \over 2}\left(z^2 - t^2\right)
\end{equation}
remains invariant.  Indeed, in Dirac's picture, the Lorentz boost is a
squeeze transformation.

\subsection{Dirac's Coupled Oscillators}

\begin{figure}
\centerline{\includegraphics[scale=1.4]{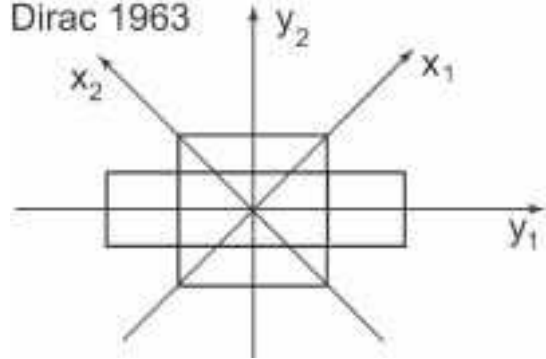}}
\caption{System of two coupled oscillators in the normal coordinate
system.  If the coupling becomes stronger, one coordinate variable
becomes contracted while the other becomes expanded.  However,
the product of the two coordinates remains constant.  This is an
area-preserving transformation in this graph, just like Lorentz-boost
in the light-cone coordinate system as described in Fig.~\ref{diracf49}.
}\label{diracf63}
\end{figure}


In 1963~\cite{dir63}, Dirac published a paper on symmetries of coupled
harmonic oscillators.  Starting from step-up and step-down operators
for the two oscillators, he was able to construct a representation of
the deSitter group $O(3,2).$  Since this group contains two $O(3,1)$
Lorentz groups, we can extract Lorentz-covariance properties from
his mathematics. It is even possible to extend this symmetry group to
$O(3,3)$ to include damping effects of the oscillators.

In the present paper, we avoid group theory and use a set of two-by-two
matrices to exploit the physical contents of Dirac's 1963 paper.
Let us start with the Hamiltonian for this system of two oscillators,
which takes the form
\begin{equation}\label{coup1}
H = {1\over 2}\left\{{1\over m}_{1}p^{2}_{1} + {1\over m}_{2}p^{2}_{2}
+ A x^{2}_{1} + B x^{2}_{2} + C x_{1} x_{2} \right\}.
\end{equation}
It is possible to diagonalize by a single rotation the quadratic
form of $x_{1}$ and $x_{2}$.  However, the momentum variables undergo
the same rotation.  Therefore, the uncoupling of the potential
energy by rotation alone will lead to a coupling of the two kinetic
energy terms.

In order to avoid this complication, we have to bring
the kinetic energy portion into a rotationally invariant form.
For this purpose, we will need the transformation
\begin{equation}\label{mtrans}
\pmatrix{p'_{1} \cr p'_{2}} = \pmatrix{ (m_{2}/m_{1})^{1/4} & 0
\cr 0 & (m_{1}/m_{2})^{1/4}} \pmatrix{p_{1} \cr p_{2}}.
\end{equation}
This transformation will change the kinetic energy portion to
\begin{equation}
{1\over 2m}\left\{p'^{2}_{1} + p'^{2}_{2} \right\}
\end{equation}
with $m = (m_{1}m_{2})^{1/2}$.
This scale transformation does not leave the $x_{1}$ and $x_{2}$
variables invariant.  If we insist on canonical
transformations~\cite{hkn95jm}, the transformation becomes
\begin{equation}
\pmatrix{x'_{1} \cr x'_{2}} = \pmatrix{ (m_{1}/m_{2})^{1/4} & 0
\cr 0 & (m_{2}/m_{1})^{1/4}} \pmatrix{x_{1} \cr x_{2}}.
\end{equation}
The scale transformations on the position variables are inversely
proportional to those of their conjugate momentum variables.  This is
based on the Hamiltonian formalism where the position and momentum
variables are independent variables.

On the other hand, in the Lagrangian formalism, where the momentum is
proportional to the velocity which is the time derivative of the
position coordinate, we have to apply the same scale transformation for
both momentum and position variables~\cite{arav89}. In this case, the
scale transformation takes the form
\begin{equation}
\pmatrix{x'_{1} \cr x'_{2}} = \pmatrix{ (m_{2}/m_{1})^{1/4} & 0
\cr 0 & (m_{1}/m_{2})^{1/4}} \pmatrix{x_{1} \cr x_{2}}.
\end{equation}
With Eq.(\ref{mtrans}) for the momentum variables, this expression does
not constitute a canonical transformation.

The canonical transformation leads to a unitary transformation in
quantum mechanics.  The issue of non-canonical transformation is not
yet completely settled in quantum mechanics and is still an open
question~\cite{hkn95jm}.  In either case, the Hamiltonian will take
the form
\begin{equation}\label{hamil2}
H = {1\over 2m}\left\{p^{2}_{1} + p^{2}_{2} \right\} +
{1\over 2}\left\{A x_{1}^{2} + B x^{2}_{2} + C x_{1} x_{2} \right\} ,
\end{equation}
Here, we have deleted for simplicity the primes on the $x$ and $p$
variables.

We are now ready to decouple this Hamiltonian by
making the coordinate rotation:
\begin{equation} \label{coup3}
\pmatrix{y_{1} \cr y_{2}} = \pmatrix{\cos\alpha & -\sin\alpha
\cr \sin\alpha & \cos\alpha} \pmatrix{x_{1} \cr x_{2}}.
\end{equation}
Under this rotation, the kinetic energy portion of the Hamiltonian in
Eq.(\ref{hamil2}) remains invariant.  Thus we can achieve the decoupling
by diagonalizing the potential energy.  Indeed, the system becomes
diagonal if the angle $\alpha$ becomes
\begin{equation}\label{coup4}
\tan 2\alpha = {C\over B - A} .
\end{equation}
This diagonalization procedure is well known.  What is new in this note
is the introduction of the new parameters $K$ and $\eta$ defined as
\begin{eqnarray}\label{coup5}
&{}& K = \sqrt{AB - C^{2}/4} ,  \nonumber \\[1.0ex]
&{}& \exp(\eta) = \frac{A + B + \sqrt{(A - B)^{2} + C^{2}}}
{\sqrt{4AB - C^{2}}} .
\end{eqnarray}

In terms of this new set of variables, the Hamiltonian can be written as
\begin{equation}\label{coup6}
H = {1\over 2m} \left\{p^{2}_{1} + p^{2}_{2} \right\} +
{K\over 2}\left\{e^{2\eta} y^{2}_{1} + e^{-2\eta} y^{2}_{2} \right\} ,
\end{equation}
with
\begin{eqnarray}\label{coup7}
&{}& y_{1} = x_{1} \cos\alpha - x_{2} \sin\alpha ,
  \nonumber \\[0.5ex]
&{}& y_{2} = x_{1} \sin\alpha + x_{2} \cos\alpha .
\end{eqnarray}
This completes the diagonalization process.  The normal frequencies
are
\begin{equation}
\omega_{1} = e^{\eta} \omega , \qquad
\omega_{2} = e^{-\eta} \omega ,
\end{equation}
with
\begin{equation}\label{omega}
\omega = \sqrt{{K \over m}} .
\end{equation}
This relatively new set of parameters has been discussed in
in connection with Feynman's rest of the universe~\cite{hkn99ajp}.

Let us go back to Eq.(\ref{hamil2}) and Eq.(\ref{coup4}).  If
$\alpha = 0$, $C$ becomes zero and the oscillators become decoupled.
If $\alpha = 45^{o}$, then $A = B$, which means that the system
consists of two identical oscillators coupled together by the
$C$ term.  In this case,
\begin{equation} \label{coup18}
\exp{(\eta)} = \sqrt{\frac{2A + C}{2A - C}}, \quad or \quad
\eta = \frac{1}{2}\ln\left(\frac{2A + C}{2A - C}\right) .
\end{equation}
Thus $\eta$ measures the strength of the coupling.

The mathematics becomes very simple for $\alpha = 45^{o}$, and this
simple case can be applied to many physical problems, including the
present problem of combining quantum mechanics with relativity.
Indeed the $y_1, y_2$ variables become
\begin{equation}
y_1 = \frac{x_1 - x_2}{\sqrt{2}}, \qquad y_2 = \frac{x_1 + x_2}{\sqrt{2}} .
\end{equation}
If $y_{1}$ and $y_{2}$ are measured in units of $(mK)^{1/4} $,
the ground-state wave function of this oscillator system is
\begin{equation}\label{coup13}
\psi_{\eta}(x_{1},x_{2}) = {1 \over \sqrt{\pi}}
\exp{\left\{-{1\over 2}(e^{\eta} y^{2}_{1} + e^{-\eta} y^{2}_{2})
\right\} } ,
\end{equation}
The wave function is separable in the $y_{1}$ and $y_{2}$ variables, when
$\eta = 0,$ remains separable while they are expanded and contracted by
$e^{\eta/2}$ and $e^{-\eta/2}$ respectively as illustrated in
Fig.~\ref{diracf63}.

On the other hand, for the variables $x_{1}$ and $x_{2}$, the wave
function takes the form
\begin{equation}
\psi_{\eta}(x_{1},x_{2}) = {1 \over \sqrt{\pi}}
\exp\left\{- {1\over 4}\left[e^{\eta}(x_{1} - x_{2})^{2} +
e^{-\eta}(x_{1} + x_{2} )^{2} \right]\right\} .
\end{equation}

In his 1963 paper~\cite{dir63}, Dirac strictly worked with step-up and
step-down operators.  He made no attempt to use a  normal coordinate
system.  It is indeed gratifying to translate his algebraic formulas
into a geometry.  Let us now compare Fig.~\ref{diracf63} with
Fig.~\ref{diracf49}.  They are the same!  Indeed, the geometry of
Lorentz boost in Dirac's light-cone coordinate system is identical
to that of the coupled oscillators.  The coupling constant is
translated into the boost parameter as given in Eq.(\ref{coup18}).

\begin{figure}
\centerline{\includegraphics[scale=1.2]{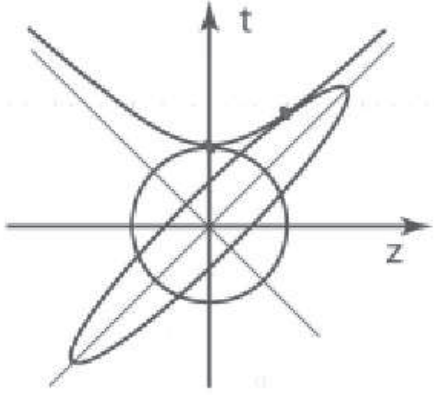}}
\caption{Effect of the Lorentz boost on the space-time
wave function.  The circular space-time distribution in the rest frame
becomes Lorentz-squeezed to become an elliptic
distribution.}\label{hyperell}
\end{figure}


\section{Lorentz-covariant Picture of Quantum Bound States}\label{dirac22}

If we combine Fig.~\ref{diracf27} and Fig.~\ref{diracf49}, then we end up
with Fig.~\ref{hyperell}.
In mathematical formula, this transformation changes the Gaussian form
of Eq.(\ref{ground}) into
\begin{equation}\label{eta}
\psi_{\eta }(z,t) = \left({1 \over \pi }\right)^{1/2}
\exp\left\{-{1\over 4}\left[e^{-\eta }(z + t)^{2} +
e^{\eta}(z - t)^{2}\right]\right\} .
\end{equation}
This formula together with Fig.~\ref{hyperell} is known to describe
all essential high-energy features observed in high-energy
laboratories~\cite{fey69b,kn77par,kn05job}.

Indeed, this elliptic deformation explains one of the most controversial
issues in high-energy physics.  Hadrons are known to be bound states of
quarks.  Its bound-state quantum mechanics is assumed to be the same as
that of the hydrogen atom.  The question is how the hadron would look
to an observer on a train.  If the train moves with a speed close to that
of light, the hadron appears like a collection of partons, according to
Feynman~\cite{fey69b}.  Feynman's partons have properties quite different
from those of the quarks.  For instance, they interact incoherently with
external signals.  The elliptic deformation property described in
Fig.~\ref{hyperell} explains that the quark and parton models are two
different manifestations of the same covariant entity.

\begin{figure}
\centerline{\includegraphics[scale=1.4]{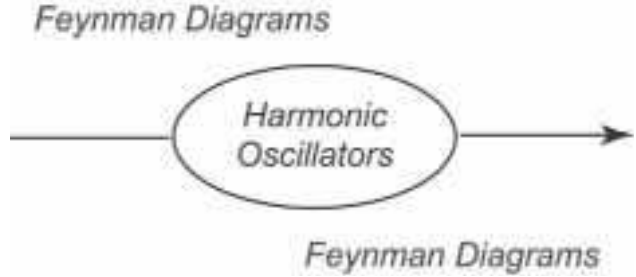}}
\vspace{5mm}
\caption{Feynman's roadmap for combining quantum mechanics with special
relativity.  Feynman diagrams work for running waves, and they provide
a satisfactory resolution for scattering states in Einstein's world.
For standing waves trapped inside an extended hadron, Feynman suggested
harmonic oscillators as the first step.}\label{dff33}
\end{figure}

Quantum field theory has been quite successful in terms of Feynman
diagrams based on the S-matrix formalism, but is useful only for physical
processes where a set of free particles becomes another set of free
particles after interaction.  Quantum field theory does not address the
question of localized probability distributions and their covariance
under Lorentz transformations.  In order to address this question,
Feynman {\it et al.} suggested harmonic oscillators to tackle the
problem~\cite{fkr71}.  Their idea is indicated in Fig.~\ref{dff33}.

In this report, we are concerned with the quantum bound system, and we
have examined the four-papers of Dirac on the question of making
the uncertainty relations consistent with special relativity.  Indeed,
Dirac discussed this fundamental problem with mathematical devices
which are both elegant and transparent.

Dirac of course noted that the time variable plays the essential
role in the Lorentz-covariant world.  On the other hand, he did
not take into consideration the concept of time separation.
When we talk about the hydrogen atom, we are concerned with the
distance between the proton and electron.  To a moving observer,
there is also a time-separation between the two particles.

Instead of the hydrogen atom, we use these days the hadron
consisting of two quarks bound together with an attractive force,
and consider their space-time positions $x_{a}$ and $x_{b}$,and
use the variables~\cite{fkr71}
\begin{equation}
X = \frac{x_{a} + x_{b}}{2} , \qquad
          x = \frac{x_{a} - x_{b}}{2\sqrt{2}} .
\end{equation}
The four-vector $X$ specifies where the hadron is located in space
and time, while the variable $x$ measures the space-time separation
between the quarks.  Let us call their time components $T$ and $t$.
These variables actively
participate in Lorentz transformations.  The existence of the
$T$ variable is known, but the Copenhagen school was not able to
see the existence of this $t$ variable.

Paul A. M. Dirac was concerned with the time variable throughout
his four papers discussed in this report.  However, he did not make
a distinction between the $T$ and $t$ variables.
The $T$ variable ranges from $-\infty$ to $+\infty$, and is constantly
increasing.  On the other hand, the $t$ variable is the time
interval, and remains unchanged in a given Lorentz frame.

Indeed, when Feynman {\it et al.} wrote down the Lorentz-invariant
differential equation~\cite{fkr71}
\begin{equation}\label{osceq}
{1\over 2} \left\{x^{2}_{\mu} -
{\partial^{2} \over \partial x_{\mu }^{2}}
\right\} \psi(x) = \lambda \psi(x) ,
\end{equation}
$x_{\mu}$ was for the space-time separation between the quarks.

This four-dimensional differential equation has more than 200
forms of solutions depending on boundary conditions.  However,
there is only one set of solutions to which we can give a physical
interpretation.  Indeed, the Gaussian form of Eq.(\ref{ground4})
is a solution of the above differential equation.  If we boost the
system along the $z$ direction, we can separate away the $x$ and
$y$ components in the Gaussian form and write the wave function
in the form of Eq.(\ref{ground}).

It is then possible to construct a representation of the
Poincar\'e group from the solutions of the above differential
equation~\cite{knp86}.  If the system is boosted, the wave function
becomes the Gaussian form given in Eq.(\ref{eta}), which becomes
Eq.(\ref{ground}) if $\eta$ becomes zero.  This wave function is
also a solution of the Lorentz-invariant differential equation of
Eq.(\ref{osceq}).  The transition from Eq.(\ref{ground}) to
Eq.(\ref{eta}) is illustrated in Fig.~\ref{hyperell}.

\section{Lorentz-covariant Quark Model}\label{quarkmo}

Early successes in the quark model includes the calculation of the
ratio of the neutron and proton magnetic moments~\cite{beg64}, and
the hadronic mass spectra~\cite{fkr71,owg67}.  These are based on
hadrons at rest.  We are interested in this paper how the hadrons
in the quark model appear to observers in different Lorentz frames.

The idea that the proton or neutron has a space-time extension had
been developed long before Gell-Mann's proposal for the quark
model~\cite{gell64}.  Yukawa~\cite{yuka53} developed this idea as
early as 1953, and his idea was followed up by Markov, Ginzburg, and
Man'ko~\cite{markov56,ginz65}.

Since Einstein formulated special relativity for point particles,
it has been and still is a challenge to formulate a theory for
particles with space-time extensions.  The most naive idea would
be to study rigid spherical objects, and there were many papers
on this subjects.  But we do not know where that story stands
these days.  We can however replace these extended rigid bodies
by extended wave packets or standing waves, thus by localized
probability entities.  Then what are the constituents within
those localized waves?  The quark model gives the natural answer
to this question.

The first experimental discovery of the non-zero size of the proton
was made by Hofstadter and McAllister~\cite{hofsta55}, who used
electron-proton scattering to measure the charge distribution inside
the proton. If the proton were a point particle, the scattering
amplitude would just be a Rutherford formula.  However, Hofstadter
and MacAllister found a tangible departure from this formula which
can only be explained by a spread-out charge distribution inside
the proton.

In this section, we are interested in how well the bound-state picture
developed in Sec.~\ref{quarkmo} works in explaining relativistic
phenomena of hadrons, specifically the proton.  The Lorentz-covariant
model of Sec.~\ref{dirac22} is of course based on Dirac's four papers
discussed in Sec.~\ref{dirac11}.

First, we show that the static quark model and Feynman's parton picture
are two limiting cases of one Lorentz-covariant entity.  In the quark
model, the hadron appears like quantum bound states with discrete
energy spectra.  In the parton model, the rapidly moving hadron appears
like a collection of infinite number of free independent particles.
Can these be explained with one theory?  This is what we would like to
address in subsection~\ref{fparton}.

As for the case between these two limits, we discuss the hadronic
form factor which occupies an important place in every theoretical
model in strong interactions since Hofstadter's discovery in
1955~\cite{hofsta55}.  The key question is the proton form factor
decreases as 1/(momentum transfer$)^4$ as the momentum transfer
becomes large.  This is called the dipole cut-off in the literature.
We shall see in subsection~\ref{formfac} that the covariant model
of Sec.~\ref{dirac22} gives this dipole cut-off for spinless quarks.

\subsection{Feynman's Parton Picture}\label{fparton}

In a hydrogen atom or a hadron consisting of two quarks, there is a
spacial separation between two constituent elements.  In the case of
the hydrogen atom we call it the Bohr radius.  If the atom or hadron is
at rest, the time-separation variable does not play any visible role
in quantum mechanics.  However, if the system is boosted to the
Lorentz frame which moves with a speed close to that of light, this
time-separation variable becomes as important as the space separation
of the Bohr radius.  Thus, the time-separation variable plays a visible role
in high-energy physics which studies fast-moving bound states.  Let
us study this problem in more detail.

It is a widely accepted view that hadrons are quantum bound states
of quarks having a localized probability distribution.  As in all
bound-state cases, this localization condition is responsible for
the existence of discrete mass spectra.  The most convincing evidence
for this bound-state picture is the hadronic mass
spectra~\cite{fkr71,knp86}.

\begin{figure}
\centerline{\includegraphics[scale=0.5]{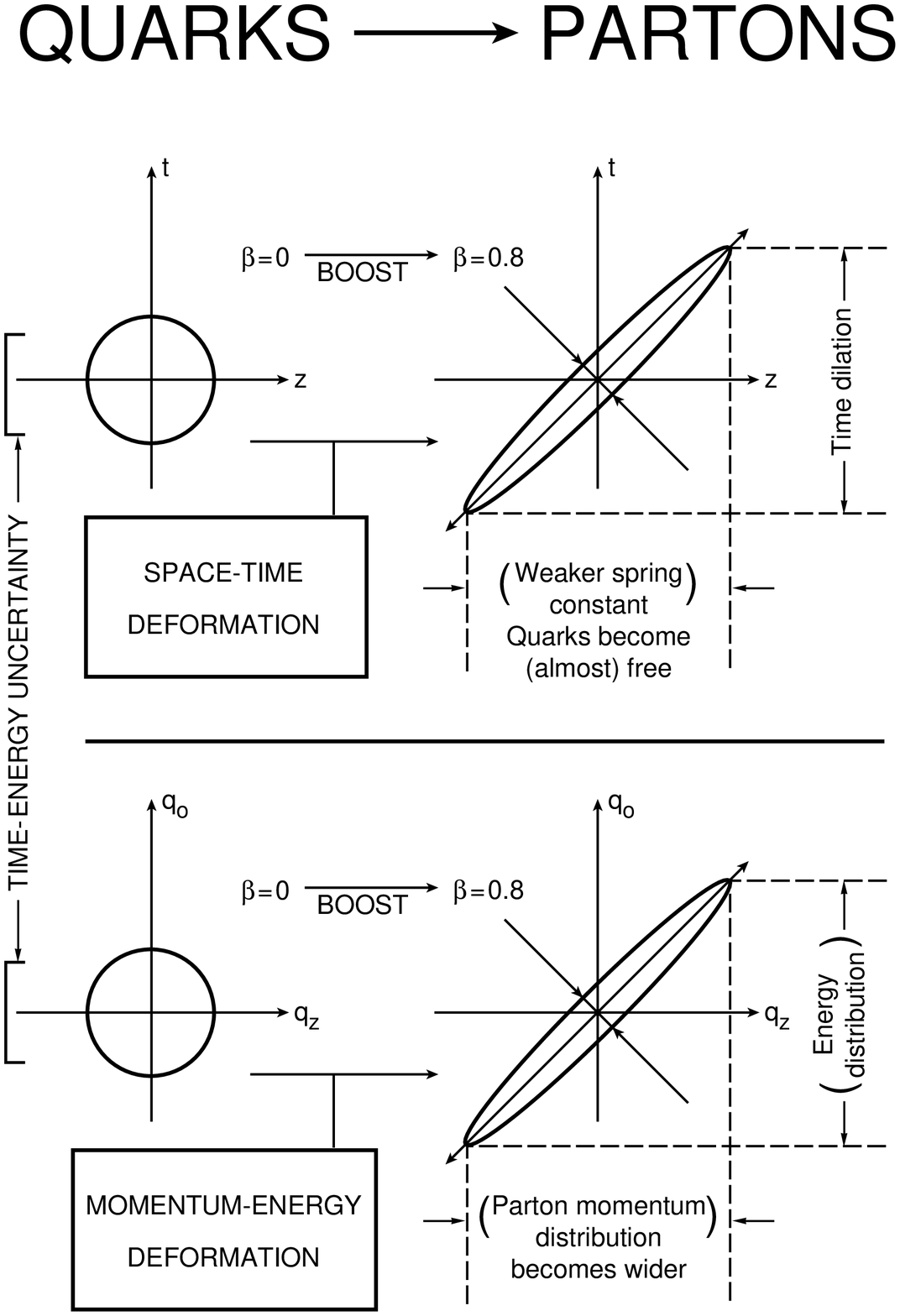}}
\vspace{5mm}
\caption{Lorentz-squeezed space-time and momentum-energy wave
functions.  As the hadron's speed approaches that of light, both
wave functions become concentrated along their respective positive
light-cone axes.  These light-cone concentrations lead to Feynman's
parton picture.}\label{parton}
\end{figure}
However, this picture of bound states is applicable only to observers
in the Lorentz frame in which the hadron is at rest.  How would the
hadrons appear to observers in other Lorentz frames?

In 1969, Feynman observed that a fast-moving hadron can be regarded
as a collection of many ``partons'' whose properties appear
to be quite different from those of the quarks~\cite{fey69b,knp86}.  For
example, the number of quarks inside a static proton is three, while
the number of partons in a rapidly moving proton appears to be infinite.
The question then is how the proton looking like a bound state of
quarks to one observer can appear different to an observer in a
different Lorentz frame?  Feynman made the following systematic
observations.

\begin{itemize}

\item[a.]  The picture is valid only for hadrons moving with
 velocity close to that of light.

\item[b.]  The interaction time between the quarks becomes dilated,
 and partons behave as free independent particles.

\item[c.]  The momentum distribution of partons becomes widespread as
 the hadron moves fast.

\item[d.]  The number of partons seems to be infinite or much larger
 than that of quarks.

\end{itemize}

\noindent Because the hadron is believed to be a bound state of two
or three quarks, each of the above phenomena appears as a paradox,
particularly b) and c) together.  How can a free particle have a
wide-spread momentum distribution?

In order to resolve this paradox, let us construct the
momentum-energy wave function corresponding to Eq.(\ref{eta}).
If the quarks have the four-momenta $p_{a}$ and $p_{b}$, we can
construct two independent four-momentum variables~\cite{fkr71}
\begin{equation}
P = p_{a} + p_{b} , \qquad q = \sqrt{2}(p_{a} - p_{b}) .
\end{equation}
The four-momentum $P$ is the total four-momentum and is thus the
hadronic four-momentum.  $q$ measures the four-momentum separation
between the quarks.  Their light-cone variables are
\begin{equation}\label{conju}
q_{u} = (q_{0} - q_{z})/\sqrt{2} ,  \qquad
q_{v} = (q_{0} + q_{z})/\sqrt{2} .
\end{equation}
The resulting momentum-energy wave function is
\begin{equation}\label{phi}
\phi_{\eta }(q_{z},q_{0}) = \left({1 \over \pi }\right)^{1/2}
\exp\left\{-{1\over 2}\left[e^{-2\eta}q_{u}^{2} +
e^{2\eta}q_{v}^{2}\right]\right\} .
\end{equation}
Because we are using here the harmonic oscillator, the mathematical
form of the above momentum-energy wave function is identical to that
of the space-time wave function of Eq.(\ref{eta}).  The Lorentz
squeeze properties of these wave functions are also the same.  This
aspect of the squeeze has been exhaustively discussed in the
literature~\cite{knp86,kn77par,kim89}, and they are illustrated again
in Fig.~\ref{parton} of the present paper.  The hadronic structure
function calculated from this formalism is in a reasonable agreement
with the experimental data~\cite{hussar81}.

When the hadron is at rest with $\eta = 0$, both wave functions
behave like those for the static bound state of quarks.  As $\eta$
increases, the wave functions become continuously squeezed until
they become concentrated along their respective positive
light-cone axes.  Let us look at the z-axis projection of the
space-time wave function.  Indeed, the width of the quark distribution
increases as the hadronic speed approaches that of the speed of
light.  The position of each quark appears widespread to the observer
in the laboratory frame, and the quarks appear like free particles.

The momentum-energy wave function is just like the space-time wave
function.  The longitudinal momentum distribution becomes wide-spread
as the hadronic speed approaches the velocity of light.  This is in
contradiction with our expectation from nonrelativistic quantum
mechanics that the width of the momentum distribution is inversely
proportional to that of the position wave function.  Our expectation
is that if the quarks are free, they must have their sharply defined
momenta, not a wide-spread distribution.

However, according to our Lorentz-squeezed space-time and
momentum-energy wave functions, the space-time width and the
momentum-energy width increase in the same direction as the hadron
is boosted.  This is of course an effect of Lorentz covariance.
This indeed leads to the resolution of one of the the quark-parton
puzzles~\cite{knp86,kn77par,kim89}.

Another puzzling problem in the parton picture is that partons appear
as incoherent particles, while quarks are coherent when the hadron
is at rest.  Does this mean that the coherence is destroyed by the
Lorentz boost?   The answer is NO, and here is the resolution to
this puzzle.

When the hadron is boosted, the hadronic matter becomes squeezed and
becomes concentrated in the elliptic region along the positive
light-cone axis.  The length of the major axis becomes expanded by
$e^{\eta}$, and the minor axis is contracted by $e^{\eta}$.

This means that the interaction time of the quarks among themselves
becomes dilated.  Because the wave function becomes wide-spread, the
distance between one end of the harmonic oscillator well and the
other end increases.  This effect, first noted by
Feynman~\cite{fey69a,fey69b},
is universally observed in high-energy hadronic experiments.  The
period of oscillation increases like $e^{\eta}$.

On the other hand, the external signal, since it is moving in the
direction opposite to the direction of the hadron, travels along
the negative light-cone axis.

If the hadron contracts along the negative light-cone axis, the
interaction time decreases by $e^{-\eta}$.  The ratio of the interaction
time to the oscillator period becomes $e^{-2\eta}$.  The energy of each
proton coming out of the Fermilab accelerator is $900 GeV$.  This leads
the ratio to $10^{-6}$.  This is indeed a small number.  The external
signal is not able to sense the interaction of the quarks among
themselves inside the hadron.

Indeed, Feynman's parton picture is one concrete physical example
where the decoherence effect is observed.  As for the entropy, the
time-separation variable belongs to the rest of the universe.  Because
we are not able to observe this variable, the entropy increases
as the hadron is boosted to exhibit the parton effect.  The
decoherence is thus accompanied by an entropy increase.

Let us go back to the coupled-oscillator system.  The light-cone
variables in Eq.(\ref{eta}) correspond to the normal coordinates in
the coupled-oscillator system given in Eq.(\ref{coup6}).  According
to Feynman's parton picture, the decoherence mechanism is determined
by the ratio of widths of the wave function along the two normal
coordinates.

This decoherence mechanism observed in Feynman's parton picture is quite
different from other decoherences discussed in the literature.  It is
widely understood that the word decoherence is the loss of coherence within
a system.  On the other hand, Feynman's decoherence discussed in this
section comes from the way the external signal interacts with the internal
constituents.

\subsection{Nucleon Form Factors}\label{formfac}

Let us first see what effect the charge distribution has on the
scattering amplitude, using nonrelativistic scattering in the Born
approximation.  If we scatter electrons from a fixed charge distribution
whose density is $e\rho(r)$, the scattering amplitude is
\begin{equation}\label{401}
f(\theta) = - \left(\frac{e^{2}m}{2\pi}\right)
 \int d^3x d^3x' \frac{\rho(r')}{R} \exp{(-i {\bf Q}\cdot {\bf x})} ,
\end{equation}
where
$r = |{\bf x}|,  R =|{\bf r} - {\bf r'} |, $ and ${\bf Q} = {\bf K_f}
- {\bf K_i},$  which is the momentum transfer.  This amplitude can
be reduced to
\begin{equation}\label{402}
f(\theta) = \frac{2me^2}{Q^2} F(Q^2) .
\end{equation}
$F(Q^2)$ is the Fourier transform of the density function which can
be written as
\begin{equation}\label{403}
F\left(Q^2\right) = \int d^3x \rho(r) \exp{(-i {\bf Q}\cdot {\bf x})} .
\end{equation}
The above quantity is called the form factor.  It describes the charge
distribution in terms of the momentum transfer.  The charge density is
normalized:
\begin{equation}\label{404}
\int \rho(r) d^3x = 1.
\end{equation}
Then F(0) = 1 from Eq.(\ref{403}).  If the density function is a delta
function corresponding to a point charge, $F\left(Q^2\right) = 1$
for all values of $Q^2,$ then the scattering amplitude of Eq.(\ref{401})
becomes the Rutherford formula for Coulomb scattering.  The deviations
from Rutherford scattering for increasing values of $Q^2$ give a measure
of the charge distribution.  This was precisely what
Hofstadter's experiment on the scattering of electrons from a proton
target found.~\cite{hofsta55}.

\begin{figure}
\centerline{\includegraphics[scale=1.0]{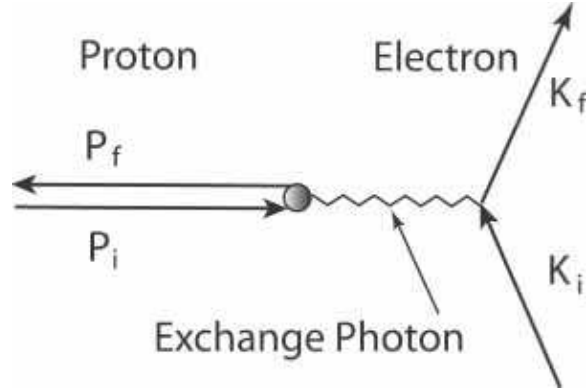}}
\vspace{5mm}
\caption{Breit frame for electron-nucleon scattering.  The momentum
of the outgoing nucleon is equal in magnitude but opposite in sign to
that of the incoming nucleon.
}\label{breit}
\end{figure}

As the energy of incoming electrons becomes higher, we have to take
into account the recoil effect of target protons, and formulate the
problem relativistically.  It is generally agreed that electrons and
their electromagnetic interaction can be described by quantum
electrodynamics, in which the method of perturbation theory using Feynman
diagrams is often employed for practical calculations~\cite{sss61,itzy90}.
In this perturbation approach, the scattering amplitude is expanded
in power series of the fine structure constant $\alpha.$  Therefore,
in lowest order in $\alpha,$ we can describe the scattering of an
electron by a proton using the diagram given in Fig.(\ref{breit}).
The corresponding matrix element is given in many textbooks on
elementary particle physics~\cite{frazer66}.  It is proportional to
\begin{equation}\label{405}
\bar{U}\left(P_f\right)\Gamma_{\mu}\left(P_f, P_i\right)
\left(\frac{1}{Q^2}\right)
\bar{U}\left(K_f\right)\gamma^{\mu}U\left(K_i\right) ,
\end{equation}
where $P_i, P_f, K_i$ and $K_f$ are the initial and final
four-momenta of the proton and electron respectively.
$U\left(P_i\right)$ is the Dirac spinor for the initial proton.
$Q^2$ is the (four-momentum transfer$)^4$ and is
\begin{equation}
      Q^2 = \left(P_f - P_i\right)^2 = \left(K_f - K_i\right)^2  .
\end{equation}
The $\left(1/Q^2\right)$ factor in Eq.(\ref{405}) comes from the
virtual photon being exchanged between the electron and the proton.
It is customary to use the letter t for $Q^2$ in form factor studies,
and this t should not be confused with the time separation variable.
In the metric we use, this quantity is positive for physical
values of the four-momenta for the particles involved in the
scattering process.

In order to make a relativistic calculation of the form factor, let
us go back to the definition of the form factor given in Eq.(\ref{403}).
The density function depends only on the target particle, and is
proportional to $\psi(x)^{\dagger}\psi(x)$, where $\psi(x)$ is the
wave function for quarks inside the proton.  This expression is a
special case of the more general form
\begin{equation}
\rho(x) = \psi_f^{\dagger}(x)\psi_i(x) ,
\end{equation}
where $\psi_i$  and  $\psi_f$ are the initial and final wave function
of the target atom.  Indeed, the form factor of Eq.(\ref{403}) can
be written as
\begin{equation}
F\left(Q^2\right) =
  \left(\psi_{f}(x), e^{-i{\bf Q}\cdot{\bf r}} \psi_{i}(x)\right) .
\end{equation}
Starting from this expression, we can make the required Lorentz
generalization using the relativistic wave functions for hadrons.

In order to see the details of the transition to relativistic physics,
we should be able to replace each quantity in the expression of
Eq.(\ref{403}) by its relativistic counterpart.  Let us go to the
Lorentz frame in which the momenta of the incoming and outgoing nucleons
have equal magnitude but opposite signs.
\begin{equation}
{\bf p_i} + {\bf p_f} = 0 .
\end{equation}
This kinematical condition is illustrated in Fig.~\ref{breit}.

The Lorentz frame in which the above condition holds is usually called
the Breit frame.  We assume without loss of generality that the proton
comes along the $z$ direction before the collision and goes out along
the negative $z$ direction after the scattering process, as illustrated
in Fig.~\ref{breit}.  In this frame, the four vector $Q = (K_f - K_i) =
(P_i - P_f)$ has no time-like component.  Thus the exponential factor
${\bf Q}\cdot {\bf r}$ can be replaced by the Lorentz-invariant form
$Q\cdot x$.  As for the wave functions for the protons, we can use
the covariant harmonic oscillator wave functions discussed in this paper
assuming that the nucleons are in the ground state.  Then the only
difference between the nonrelativistic and relativistic cases is that
the integral in the evaluation of Eq.(\ref{403}) is four-dimensional,
including that for the time-like direction.  This integral in the
time-separation variable does not interfere with the exponential
factor which does not depend on the time-separation variable.

Let us now write down the integral:
\begin{equation}\label{406}
g\left(Q^2\right) = \int d^4x \psi^{\dagger}_{-\beta}(x)
  \psi_{\beta}(x) \exp{(-iQ\cdot x)} .
\end{equation}
where $\beta$ is the velocity parameter for the incoming proton, and
the wave function $\psi_{\beta}$ takes the form:
\begin{eqnarray}
&{}&\psi_{\beta}(x) = \frac{1}{\pi}
       \exp{\left\{ - \frac{x^2 + y^2}{2}\right\}}  \nonumber\\[2ex]
&{}& \hspace{10mm} \times
\exp{\left\{ -\frac{1}{4}\left[\frac{1 - \beta}{1 + \beta} (z + t)^2
 + \frac{1 + \beta}{1 - \beta}(z - t)^2 \right]
 \right\}} .
\end{eqnarray}

After the above decomposition of the wave functions, we can perform
the integrations in the $x$ and $y$ variables trivially.  After
dropping these trivial factors, we can write the product of the two
wave functions as
\begin{equation}\label{408}
\psi^{\dagger}_{-\beta}(x)\psi_{\beta}(x)
         = \frac{1}{\pi} \exp{\left\{-\left(\frac{1 + \beta^2}
         {1 - \beta^2}\right) \left(t^2 + z^2\right)\right\}} .
\end{equation}
Thus the $z$ and $t$ variables have been separated.  Since the exponential
factor in Eq.(\ref{403}) does not depend on $t$, the $t$ integral in
Eq.(\ref{406}) can also be trivially performed, and the integral
of Eq.(\ref{406}) can be written as
\begin{equation}\label{407}
g\left(Q^2 \right) = \sqrt{\frac{1}{\pi}\left(\frac{1 - \beta^2}
  {1 + \beta^2}\right)} \int dz e^{-2iPz} \exp{\left\{- \frac{1 + \beta^2}
  {1 - \beta^2} z^2 \right\}} ,
\end{equation}
where $P$ is the $z$ component of the momentum of the incoming nucleon.
The (momentum transfer$)^2$ variable $Q^2$ is $4P^2.$  Indeed, the
distribution of the hadronic material along the longitudinal direction
became contracted~\cite{licht70}.

We note that $\beta$ can be written as
\begin{equation}
 \beta^2 = \frac{Q^2}{Q^2 + 4M^2} ,
\end{equation}
where $M$ is the proton mass.  This equation tells $\beta = 0$ when
$Q^2 = 0$, while it becomes one as $Q^2$ becomes infinity.

The evaluation of the above integral for $g\left(Q^2\right)$ in
Eq.(\ref{407}) leads to
\begin{equation}
 g\left(Q^2\right) = \left(\frac{2M^2}{Q^2 + 2M^2}\right)
 \exp{\left\{\frac{-Q^2}{2(Q^2 + 2M^2)}\right\}} .
\end{equation}
For $Q^2 = 0,$ the above expression becomes 1.  It decreases as
\begin{equation}\label{409}
 g\left(Q^2\right) \sim  \frac{1}{Q^2}
\end{equation}
for large values of $Q^2.$

We have so far carried out the calculation for an oscillator bound
state of two quarks.  The proton consists of three quarks.  As shown
in the paper of Feynman {\it et al.}~\cite{fkr71}, the problem becomes
a product of two oscillator modes. Thus, the generalization of the
above calculation to the three-quark system is straightforward, and
the result is that the form factor $G\left(Q^2\right)$ takes the form
\begin{equation}
 G\left(Q^2\right) = \left(\frac{2M^2}{Q^2 + 2M^2}\right)^2
 \exp{\left\{\frac{- Q^2}{\left(Q^2 + 2M^2\right)}\right\}} .
\end{equation}
which is 1 at $Q^2 = 0$, and decreases as
\begin{equation}\label{410}
 G\left(Q^2\right) \sim \left[\frac{1}{Q^2}\right]^2
\end{equation}
for large values of $Q^2.$  Indeed, this function satisfies the
requirement of the ``dipole-cut-off'' behavior for of the form
factors, which has been observed in high-energy laboratories.
This calculation was carried first by Fujimura {\it et al.} in
1970~\cite{fuji70}.

Let us re-examine the above calculation.  If we replace $\beta$
by zero in Eq.(\ref{407}) and ignore the elliptic deformation of the
wave functions, $g\left(Q^2\right)$ will become
\begin{equation}
g\left(Q^2\right) = e^{-Q^2/4} ,
\end{equation}
which will leads to an exponential cut-off of the form factor.  This
is not what we observe in laboratories.

In order to gain a deeper understanding of the above-mentioned
correlation, let us study the case using the momentum-energy wave
functions:
\begin{equation}
 \phi_{\beta}(q) = \left(\frac{1}{\pi}\right)^2
 \int d^4 e^{-iq\cdot x} \psi_\beta (x) .
\end{equation}

As before, we can ignore the transverse components.  Then
$g\left(Q^2\right)$ can be written as~\cite{knp86}
\begin{equation}
 \int  dq_0 dq_z \phi^*_\beta \left(q_0, q_z - P\right)
 \phi_\beta \left(q_0, q_z + P \right) .
\end{equation}

\begin{figure}
\centerline{\includegraphics[scale=1.4]{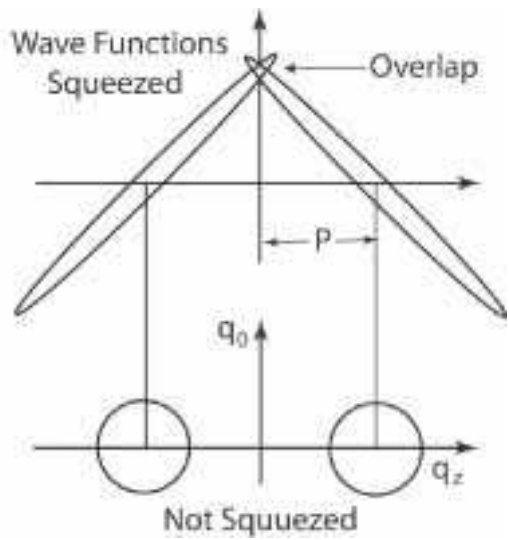}}
\vspace{5mm}
\caption{ Lorentz-Dirac deformation of the momentum-energy wave
functions in the form factor calculation.  As the momentum transfer
increases, the two wave functions become separated.  In the relativistic
case, the wave functions maintain an overlapping region.  Wave functions
become completely separated in the nonrelativistic calculation.  This
lack of overlapping region leads to an unacceptable behavior of the form
factor.}\label{overlap}
\end{figure}

We have sketched the above overlap integral in Fig.~\ref{overlap}.
When $Q^2 = 0$ or $P = 0$, the two wave functions overlap completely
in the $q_z q_0$ plane.  As P increases, the wave functions become
separated.  However, they maintain a small overlapping region due
to the elliptic or squeeze deformation seen in Fig.~\ref{hyperell}.
In the non-relativistic case, where the deformation is not taken
into account, there is no overlapping region as seen in Fig.~\ref{overlap}.
This is precisely why the relativistic calculation gives a slower
decrease in $Q^2$ than in the nonrelativistic case.

We have so far been interested only in the space-time behavior of
the hadronic wave function.  We should not forget the fact that quarks
are spin-1/2 particles.  The effect of this spin manifests itself
prominently in the baryonic mass spectra.  Since we are concerned here
with the relativistic effects, we have to construct a relativistic spin
wave function for the quarks.  This quark wave function should give the
hadronic spin wave function.  In the case of nucleons, the quark spins
should be combined in a manner to generate the form factor of
Eq.(\ref{408}).

The naive approach to this problem is to use free Dirac spinors
for the quarks.  However, it was shown by Lipes~\cite{lipes72} that
the use of free-particle Dirac spinors leads to a wrong form
factor behavior.  Since quarks in a hadron are not free particles,
Lipes's result does not alarm us.  The difficult problem is to find a
suitable mechanism in which quark spins are coupled to orbital motion in
a relativistic manner.  This is a nontrivial research problem, and
further study is needed along this direction~\cite{henriq75}.

In addition, there are recent experimental results which indicate
departure from the dipole behavior of Eq.(\ref{410})~\cite{punjabi05}.
In addition, there have been other theoretical attempts to calculate
the proton form factor.  Yes, whenever a new theoretical model appears,
there appears a new attempt to calculate the form factor.  In the past,
there were many attempts to calculate this quantity in the framework
of quantum field theory, without much success.

In 1960, Frazer and Fulco calculated the form factor using the
technique of dispersion relations~\cite{frazer60}.  In so doing
they had to assume the existence of the so-called $\rho$ meson,
which was later found experimentally, and which subsequently played
a pivotal role in the development of the quark model.

Even these days, the form factor calculation occupies a very important
place in recent theoretical models, such as QCD lattice
theory~\cite{matevo05} and the Faddeev equation~\cite{roberts05}.
However, it is still noteworthy that Dirac's form of Lorentz-covariant
bound states leads to the essential dipole cut-off behavior of the
proton form factor.

\section*{Conclusion}
The hydrogen atom played a pivotal role in the development of quantum
mechanics.  Quantum mechanics had to be formulated to explain its
discrete energy spectra, radiation decay rates, as well as the scattering
of electrons by the proton.

The quark model still plays the central role in present-day high-energy
physics.  The model can explain the hadronc mass spectra, hadronic
structure, and hadronic decay rates.  In addition we are dealing with
hadrons which appear as quantum bound states when they are at rest, how
would they appear to observes in different frames, particularly in the
frame moving with a velocity close to that of light.  It was Feynman
who proposed the parton model to describe those high-energy hadrons.

In this model, we have presented a Lorentz-covariant model which gives
both the quark model and the parton model as two limiting cases.  In
addition we discussed the form factor as the case between these
limits.

In constructing the Lorentz-covariant model, we noted Paul A. M.
Dirac made life-long efforts to make quantum mechanics consistent
with special relativity.  We have chosen four of his papers and
combined them to construct a consistent theory.  It was like building
a canal.  The easiest way to build the canal is to link up the
existing lakes.  Dirac indeed dug four big lakes.  It is a
gratifying experience to link them up.

Dirac constructed those lakes in order to study
whether the Copenhagen school of quantum mechanics can be made
consistent with Einstein's Lorentz-covariant world.

After studying Dirac's papers, we arrived at the conclusion that
the Copenhagen school completely forgot to take into account
the question of simultaneity and time separation~\cite{kn06aip}.
The question then is whether the localized probability distribution
can be made consistent with Einstein's Lorentz covariance.  We
have addressed this question in this paper.

\section*{ACKNOWLEDGMENT}

We would like to thank Stephen Wallace for telling us about both
experimental and theoretical aspects of form factor studies.

\end{document}